\documentclass{PoS}
\usepackage{cite}
\usepackage{amsmath}
\usepackage{amssymb}
\usepackage{epsfig}
\usepackage{etex, xy}
\xyoption{all}
\newcommand{\gsim}{\raisebox{-0.07cm   }
{$\, \stackrel{>}{{\scriptstyle\sim}}\, $}}
\newcommand{\GeV}{{\rm GeV}}

\usepackage{tikz}
\usetikzlibrary{matrix}

\allowdisplaybreaks[4]

\title{{\rm \footnotesize DESY 16-149,~~DO-TH 16/16,~~MITP/16-096}\\
New Results on Massive 3-Loop Wilson Coefficients in Deep-Inelastic scattering\thanks{This work 
was supported in part by the Austrian Science Fund (FWF) grant SFB F50 (F5009-N15), the European 
Commission through contract PITN-GA-2012-316704 ({HIGGSTOOLS}) and by FP7 ERC Starting Grant  257638 PAGAP.}
\footnote{Based on the talks presented by A.~Behring, A. De Freitas, and G.~Falcioni.}
\footnote{Also contribution to the Proceedings of DIS 2016.}
}

\ShortTitle{New Results on Massive 3-Loop Wilson Coefficients in Deep-Inelastic Scattering}

\author{J.~Ablinger\\
Research Institute for Symbolic Computation (RISC),
                          Johannes Kepler University, Altenbergerstra\ss{}e 69,
                          A--4040, Linz, Austria}

\author{A.~ Behring, J.~Bl\"umlein, G.~Falcioni, A.~De Freitas\\
       Deutsches Elektronen-Synchrotron, DESY, Platanenallee 6, D-15738 Zeuthen, Germany}

\author{A.~Hasselhuhn\\
Institut f\"ur Theoretische Teilchenphysik
Campus S\"ud, Karlsruher Institut f\"ur Technologie (KIT), D-76128 Karlsruhe, Germany}

\author{A.~von Manteuffel\\
Department of Physics and Astronomy, Michigan State University, East Lansing, MI 48824, USA\\
PRISMA Cluster of Excellence, Johannes Gutenberg University, 55099 Mainz, Germany
}

\author{M.~Round, C.~Schneider\\
Research Institute for Symbolic Computation (RISC),
                          Johannes Kepler University, Altenbergerstra\ss{}e 69,
                          A--4040, Linz, Austria}

\author{F.~Wi\ss{}brock\\
IHES, 35 Route de Chartres, F-91440 Bures-sur-Yvette, France}

\abstract{We present recent results on newly calculated 2- and 3-loop contributions to the heavy quark 
parts of the structure functions in deep-inelastic scattering due to charm and bottom.}

\FullConference{Loops and Legs in Quantum Field Theory\\
                 24-29 April 2016\\
                 Leipzig, Germany}

\makeatletter
\g@addto@macro\bfseries{\boldmath}
\makeatother
\begin{document}
%-------------------------------------------------------------------------------------------------
\section{Introduction}
\label{sec:1}
%-------------------------------------------------------------------------------------------------

\vspace*{1mm}
\noindent
At present the 3-loop heavy flavor corrections form the missing link to analyze the World deep-inelastic scattering
data at next-to-next-to leading order with respect to the determination of the unpolarized parton distribution functions (PDFs) 
\cite{PDF},
the strong coupling constant \cite{ALPHA}, and the masses of the charm and bottom quarks \cite{MCMB}. These 
distributions and parameters form an essential input for all precision measurements at the LHC, notably the measurement of
the Higgs boson \cite{HIGGS} and $t\bar{t}$ \cite{TTBAR} production cross sections. Furthermore, their precise knowledge 
provides new unprecedented tests of the Standard Model and may help in this way to reveal potential deviations pointing to
new physics. 

The present project is devoted to calculate the 3-loop heavy flavor Wilson coefficients at large scales $Q^2 \gg m^2$ in analytic
form. In case of the structure function $F_2(x,Q^2)$ this approximation suffices at the $1\%$ level for scales $Q^2/m^2 \gsim 
10$ \cite{Buza:1995ie}, a region to be selected anyway in case of charm to stay off  higher twist terms \cite{Alekhin:2012ig} at 
this precision. 

In 2009 a series of Mellin moments has been calculated for the corresponding massive operator matrix elements (OMEs) ranging 
from $N=2$ to $N=10...14$, depending on the Wilson coefficient, in Refs.~\cite{Bierenbaum:2009mv} by mapping the operator matrix 
elements for the different moments onto massive tadpoles, which were computed using the code {\tt MATAD} \cite{Steinhauser:2000ry}.
These moments serve now 
for comparison in the computation of the general-$N$ results. In the time since, four out of five neutral current massive 3-loop 
Wilson coefficients have been calculated \cite{Ablinger:2010ty,Behring:2014eya,Ablinger:2014vwa,Ablinger:2014nga,Blumlein:2006mh}, 
along with seven out of eight massive OMEs 
\cite{Ablinger:2010ty,Behring:2014eya,Ablinger:2014vwa,Ablinger:2014nga,Blumlein:2012vq,Ablinger:2014uka,Ablinger:2014lka,
Ablinger:2016eyd,AGG}. For all processes the logarithmic 3-loop contributions are known \cite{Behring:2014eya}. To perform these 
computations, several technical, 
computer-algebraic and mathematical innovations were necessary, which are described in 
Refs.~\cite{Blumlein:1998if,Vermaseren:1998uu,Blumlein:2003gb,Blumlein:2009ta,Blumlein:2009fz,Blumlein:2009cf,%%
Blumlein:2009tj,Blumlein:2010zv,Moch:2001zr,Ablinger:2013cf,Ablinger:2014bra,Ablinger:2011te,
Ablinger:2012qm,Ablinger:2014yaa,Ablinger:2015tua,
Ablinger:2016yjz,SURV}. The different physics results, including also necessary 2-loop calculations, for the neutral 
and 
charged current reactions  were published in 
Refs.~\cite{Bierenbaum:2007qe,Bierenbaum:2008yu,Bierenbaum:2009zt,Blumlein:2014fqa,Blumlein:2016xcy} 
and recent surveys were given in \cite{Blumlein:2014zxa,Ablinger:2016eyd}.

In this note we describe recent developments of the project. The paper is organized as follows. The basic formalism and 
a series of technical details of the calculation are summarized in Section~\ref{sec:2}. In Section~\ref{sec:3} we review the
status of the calculation of the neutral current structure function $F_2(x,Q^2)$. A survey on the calculation of the 
3-loop 
corrections to non-singlet charged and neutral current structure functions is given in Section~\ref{sec:4}. Results on the
calculation of the heavy flavor corrections in the full region of $Q^2$ for various  polarized and unpolarized non-singlet structure 
functions to $O(\alpha_s^2)$ and associated sum rules are given in Section~\ref{sec:5}. Recent results of 3-loop two-mass corrections
are reported in Section~\ref{sec:6}, and Section~\ref{sec:7} contains the conclusions.

\vspace*{-2mm}
%-------------------------------------------------------------------------------------------------
\section{Basic Formalism and Technical Aspects of the Calculation}
\label{sec:2}
%-------------------------------------------------------------------------------------------------

\vspace*{1mm}
\noindent
As it has been outlined in Refs.~\cite{Bierenbaum:2009mv,Behring:2014eya} the massive 3-loop Wilson coefficients for 
deep-inelastic scattering can be represented in terms of the massless Wilson coefficients \cite{ZN,Vermaseren:2005qc} and 
massive operator matrix elements in the asymptotic region $Q^2 \gg m^2$, if the single heavy quark case is considered.
Here $Q^2$ denotes the virtuality of the process and $m$ the heavy quark mass. In the present project the massive
OMEs are newly computed and used in the calculation of the corresponding massive Wilson coefficients. Furthermore, the massive OMEs
describe the transition of a single heavy flavor becoming light in the variable flavor number scheme (VFNS), 
cf.~\cite{Buza:1996wv,Behring:2014eya}. Up to 2-loop order the Wilson coefficients are know (semi)analytically 
\cite{Laenen:1992zk}\footnote{For a precise implementation in Mellin space, see \cite{Alekhin:2003ev}.}. In the asymptotic region 
analytic 
expressions have been derived in Refs.~\cite{Buza:1995ie,Bierenbaum:2007qe}. 

The whole 3-loop project constitutes of a calculation of 2864 Feynman diagrams mapping to some dozens of $10^4$ scalar integrals.
They are generated using {\tt QGRAF} \cite{QGRAF}, calculating their Dirac structure using {\tt FORM} \cite{FORM} and the color 
structure using {\tt Color} \cite{vanRitbergen:1998pn}.
Except for simpler topologies, we use the integration by part (IBP)-reduction \cite{IBP} implemented in the package {\tt Reduze\;2} 
\cite{REDUZE}\footnote{The package uses the codes {\tt GiNaC} \cite{Bauer:2000cp} and {\tt Fermat} \cite{FERMAT}.} to map
the major part of the problem to the calculation of 687 master integrals, out of which 116 remain to be calculated. 571 master integrals 
can be solved by sum-structures using difference-field theory 
\cite{Karr:81,Schneider:01,Schneider:05a,Schneider:07d,Schneider:10b,Schneider:10c,Schneider:15a,Schneider:08c,DFTheory} implemented in 
the 
package {\tt Sigma} \cite{SIG1,SIG2}, or other techniques, like  the use of hypergeometric functions \cite{HYP}, Mellin-Barnes 
representations \cite{MB}, the method of hyperlogarithms \cite{HYPLOG,Ablinger:2014yaa}, the solution of differential 
equations \cite{DEQ}, and the Almkvist-Zeilberger algorithm \cite{AZ},
implemented within the package {\tt MultiIntegrate} \cite{Ablinger:PhDThesis}.

They can be expressed by iterative sum-structures in $N$-space or iterative integrals in $x$-space, i.e. in terms of harmonic 
polylogarithms
\cite{Remiddi:1999ew}, generalized harmonic polylogarithms of the Kummer-type \cite{Moch:2001zr,Ablinger:2013cf}, cyclotomic harmonic 
polylogarithms \cite{Ablinger:2011te}, or root-valued
iterated integrals \cite{Ablinger:2014bra}, which all correspond to difference or differential equations that can be factorized to 
first order.
This is not the case for all the remaining 116 master integrals, which depend on structures obeying 2nd order equations. Here we 
expect complete elliptic integrals of rational argument  and related functions to emerge, see \cite{ELLIPT}\footnote{Similar structures 
were observed also in Refs.~\cite{ERLT,Adams:2016xah,Adams:2015ydq}.}, over which first order structures are iterated again. We 
currently work on the solution of these systems.

In all these methods the solution of recurrences and the various properties of special functions emerging in this context play a central 
role and have to be used algorithmically. This is made possible by the packages {\tt Sigma} \cite{SIG1,SIG2}, {\tt EvaluateMultiSums} and 
{\tt SumProduction}, \cite{EMSSP}, {\tt RhoSum} \cite{RHOSUM}, decoupling formalisms \cite{DECOUP}, {\tt 
HarmonicSums} \cite{Ablinger:PhDThesis,Harmonicsums,Ablinger:2013cf,Ablinger:2011te,Ablinger:2014bra}
and {\tt MultiIntegr- ate} \cite{Ablinger:PhDThesis}.

\vspace*{-2mm}
%-------------------------------------------------------------------------------------------------
\section{Status of the 3-Loop Neutral Current Corrections}
\label{sec:3}
%-------------------------------------------------------------------------------------------------

\vspace*{1mm}
\noindent
The heavy flavor contributions to the structure functions $F_{2,L}(x,Q^2)$ are given by 
%------------------------------------------------------------------------------------------
\begin{eqnarray}
F_{2,L}^{\rm heavy}(x,Q^2) =F_{2,L}(x,Q^2) - F_{2,L}^{\rm massless}(x,Q^2)~,
\end{eqnarray}
%------------------------------------------------------------------------------------------
for a single heavy quark $Q$ and $N_F$ massless quarks
%------------------------------------------------------------------------------------------
\begin{eqnarray}
\label{eqF2}
F_{(2,L)}^{\rm heavy}(x,N_F\!\!\!&+&\!\!\!1,Q^2,m^2) =\nonumber\\
       &&\hspace*{-1cm} \sum_{k=1}^{N_F}e_k^2\Biggl\{
                   L_{q,(2,L)}^{\sf NS}\left(x,N_F+1,\frac{Q^2}{m^2}
                                                ,\frac{m^2}{\mu^2}\right)
                \otimes
                   \Bigl[f_k(x,\mu^2,N_F)+f_{\overline{k}}(x,\mu^2,N_F)\Bigr]
\nonumber\\ &&\hspace{4mm}
               +\frac{1}{N_F}L_{q,(2,L)}^{\sf PS}\left(x,N_F+1,\frac{Q^2}{m^2}
                                                ,\frac{m^2}{\mu^2}\right)
                \otimes
                   \Sigma(x,\mu^2,N_F)
\nonumber\\ &&\hspace{4mm}
               +\frac{1}{N_F}L_{g,(2,L)}^{\sf S}\left(x,N_F+1,\frac{Q^2}{m^2}
                                                 ,\frac{m^2}{\mu^2}\right)
                \otimes
                   G(x,\mu^2,N_F)
                             \Biggr\}
\nonumber\\
       &+&  e_Q^2\Biggl[
                   H_{q,(2,L)}^{\sf PS}\left(x,N_F+1,\frac{Q^2}{m^2}
                                        ,\frac{m^2}{\mu^2}\right)
                \otimes
                   \Sigma(x,\mu^2,N_F)
\nonumber\\ && \hspace{7mm}
                  +H_{g,(2,L)}^{\sf S}\left(x,N_F+1,\frac{Q^2}{m^2}
                                           ,\frac{m^2}{\mu^2}\right)
                \otimes
                   G(x,\mu^2,N_F)
                                  \Biggr]~.
\end{eqnarray}
%-------------------------------------------------------------------------------------------------
Here $e_k$ and $e_Q$ denote the light and heavy quark charges, $f_k$ and $G$ are the quarkonic and gluon
parton densities and $\Sigma = \sum_{k=1}^{N_F} \left(f_k + f_{\bar{k}}\right)$ the singlet density, 
and $\mu$ denotes the factorization  
scale. The five heavy flavor Wilson coefficients are given by $L_{q,}^{\sf NS}, L_{q}^{\sf PS}, L_{g}^{\sf S}$ and 
$H_q^{\sf PS}, H_g^{\sf S}$, cf.~Refs.~\cite{Ablinger:2010ty,Behring:2014eya,Ablinger:2014vwa,Ablinger:2014nga}.

\begin{figure}\centering
     \includegraphics[width=0.7\textwidth]{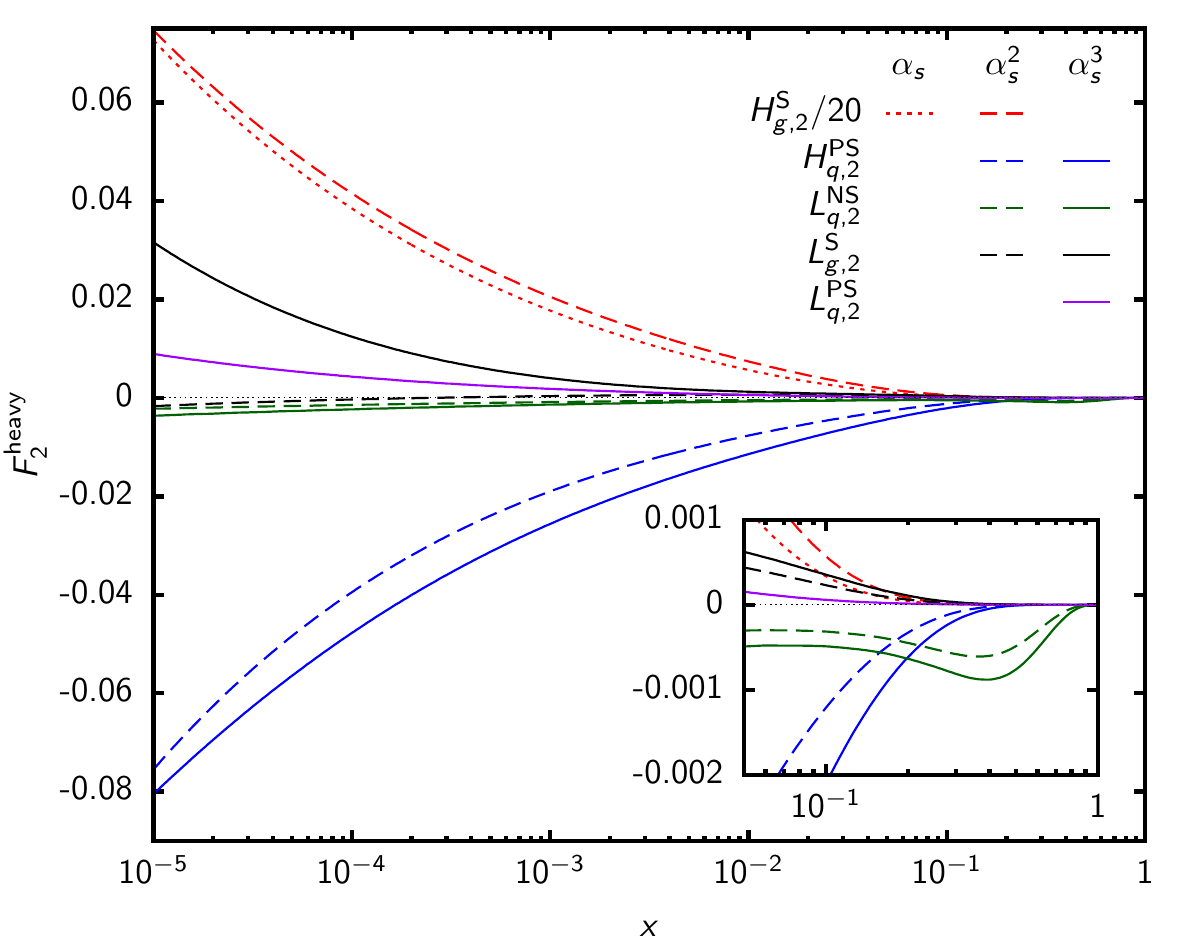}
\caption[]{The different known contributions to $F_{2}^{\rm charm}(x,Q^2)$ by the five heavy flavor Wilson coefficients
from $O(\alpha_s)$ (dotted lines),  $O(\alpha_s^2)$ (dashed lines), $O(\alpha_s^2)$ (full lines),
for $m_c^{\rm pole} = 1.59~\GeV$ and the PDFs \cite{Alekhin:2013nda}. The contributions due to $H_g^{\sf S}$ have been scaled down by 
a factor of 20 for better visibility.
\label{fig:1}
}
\end{figure}

In Figure~\ref{fig:1} the present status of the 3-loop results on $F_2^{\rm charm}(x,Q^2)$ is illustrated for $Q^2 = 100~\GeV^2$
referring to the parton densities \cite{Alekhin:2013nda} and $m_c^{\rm pole} = 1.59~\GeV$. We show the different contributions 
in $O(\alpha_s), O(\alpha_s^2)$ and $O(\alpha_s^3)$ using the asymptotic representation,
highlighting their behaviour at large values of $x$ in the inset.

\noindent
%-------------------------------------------------------------------------------------------------
%-------------------------------------------------------------------------------------------------

\vspace*{-3mm}
The intended accuracy is $O(1\%)$ and requires all the contributions mentioned above, at least in some part of the 
kinematic range.
The largest contribution is due to $H_g^{\sf S}$ driven by the gluon distribution starting at $O(\alpha_s)$. Here the calculation of the 
3-loop term is underway, while all other contributions are known already. The next largest term in the small $x$ region is  
$H_q^{\sf 
PS}$, yielding negative corrections there. It is followed by $L_g^{\sf S}$ with smaller positive corrections at 3-loop order, which are
larger than those in 2-loop order by the same Wilson coefficient. The fourth largest contribution is due to $L_q^{\sf PS}$ and 
the smallest contributions are due to $L_q^{\sf NS}$, while it is largest in the valence region. The full knowledge of the 3-loop heavy 
flavor corrections to $F_2(x,Q^2)$ will
improve on presently remaining theory errors both on $\alpha_s(M_Z^2)$ and $m_c$ \cite{MCMB} due to a yet approximate treatment 
\cite{Kawamura:2012cr} of these corrections, based on the previous work 
\cite{Bierenbaum:2009mv,Bierenbaum:2007qe,Bierenbaum:2008yu,Ablinger:2010ty,Bierenbaum:2009zt}.

\vspace*{-2mm}
%-------------------------------------------------------------------------------------------------
\section{3-Loop Non-Singlet Polarized and Charged Current Corrections}
\label{sec:4}
%-------------------------------------------------------------------------------------------------

\vspace*{1mm}
\noindent
The flavor non-singlet OME \cite{Ablinger:2014vwa} appears in various unpolarized and polarized neutral 
and charged current structure functions, combined with the different massless Wilson coefficients and helicity
dependent anomalous dimensions \cite{Vermaseren:2005qc,Moch:2008fj,Davies:2016ruz,Moch:2014sna}. This allows to
calculate the polarized 3-loop massive non-singlet Wilson coefficient for twist-2 part of the structure
functions $g_{1,2}(x,Q^2)$ and the charged current structure functions $xF_3^{W^+ - W^-}(x,Q^2)$ and 
$F_{1,2}^{W^+ - W^-}(x,Q^2)$, cf.~Refs.~\cite{Behring:2015zaa,Behring:2015roa,F1F2}; for the lower order corrections
see \cite{Blumlein:2014fqa,CC}. Here, $g_2$ is obtained
by the Wandzura-Wilczek relation \cite{Wandzura:1977qf,Jackson:1989ph,Blumlein:1996vs,Blumlein:1998nv}.

\begin{figure}\centering
\includegraphics[width=0.49\textwidth]{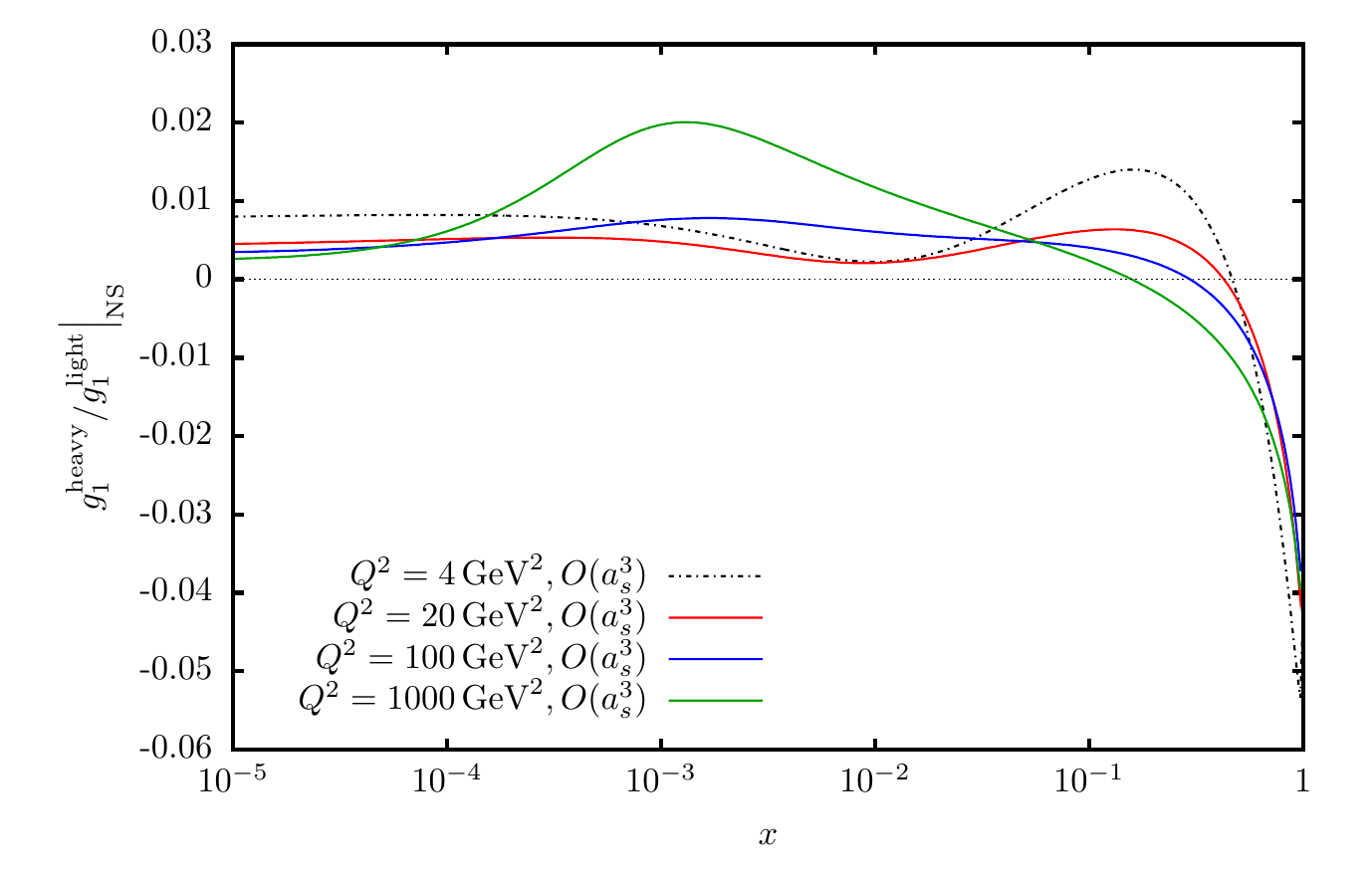}
\includegraphics[width=0.49\textwidth]{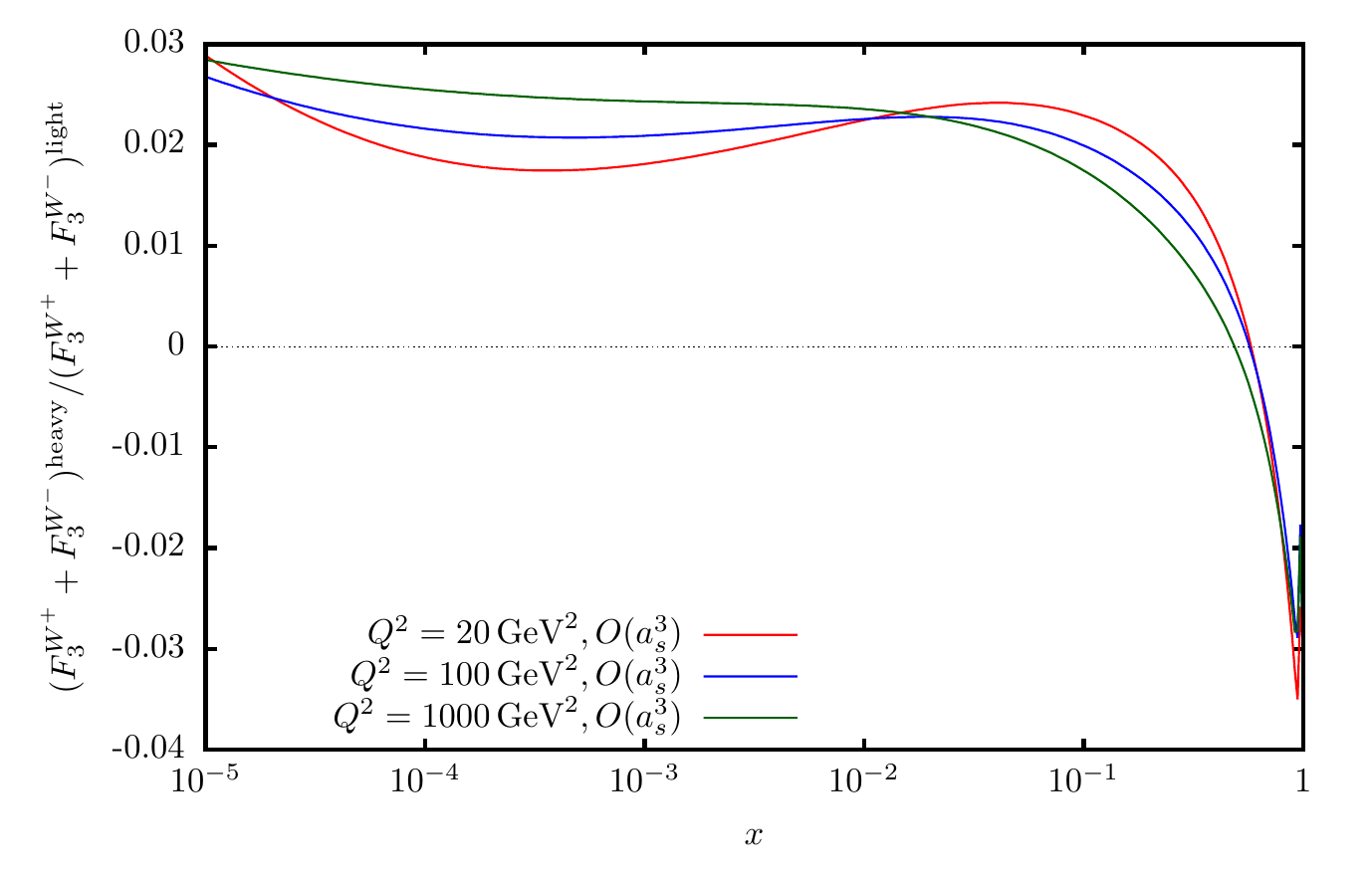}
\caption[]{The relative 3-loop charm quark corrections to the polarized neutral current structure function
$g_1^{\rm NS}(x,Q^2)$ using the PDFs \cite{Blumlein:2010rn} and the unpolarized charged current 
structure function $xF_3^{W^+ + W^-}(x,Q^2)$ using the 
PDFs \cite{Alekhin:2013nda} for $m_c^{\rm pole} = 1.59~\GeV$ as functions of $x$ and $Q^2$; 
from \cite{Behring:2015zaa} and \cite{Behring:2015roa}.
\label{fig:2}}
\end{figure}

In Figure~\ref{fig:2} we show the relative 3-loop corrections to $g_{1}^{\rm NS}(x,Q^2)$ and $xF_3^{W^+ + 
W^-}(x,Q^2)$ for typical scales $Q^2$ using the polarized PDFs \cite{Blumlein:2010rn} and the
unpolarized PDFs \cite{Alekhin:2013nda}, normalizing to the massless contributions.
\noindent
%-------------------------------------------------------------------------------------------------
%-------------------------------------------------------------------------------------------------

\vspace*{-3mm}
The 3-loop heavy flavor corrections to $g_1^{\rm NS}$ vary between $+2\%$ and $-5\%$ and those 
for the charged current structure function $xF_3^{W^+ + W^-}$ between $+3\%$ and $-3\%$, compared to
the light flavor contributions. These corrections cannot be resolved in present measurements but 
will play a role in high luminosity measurements at planned future colliders \cite{FUTURE}.

The asymptotic 3-loop heavy flavor corrections to the charged current structure functions $F_{1(2)}^{W^+ - W^-}$
is shown in Figure~\ref{fig:3}. The relative contribution ranges from $-1 (-2)\%$ to $-8\%$, with the largest corrections
around $x \sim 0.03$. The higher order terms yield positive corrections.

\vspace*{-2mm}
%-------------------------------------------------------------------------------------------------
\section[]{\boldmath Power Corrections in the Non-Singlet Case at 
\boldmath {\boldmath $\mathbf{O(\alpha_s^2)}$}}
\label{sec:5}
%-------------------------------------------------------------------------------------------------

\vspace*{1mm}
\noindent
In the foregoing sections we have presented results calculating the massive Wilson coefficients in the asymptotic region
$Q^2 \gg m^2$. It is, however, also interesting to study the Wilson coefficients 
at lower scales $Q^2$ for data in this
region and to estimate the scales at which the asymptotic representation is valid. We investigate the non-singlet 
massive Wilson coefficients to $O(\alpha_s^2)$ for the inclusive structure functions \cite{Blumlein:2016xcy}.

\noindent
%-------------------------------------------------------------------------------------------------
\begin{figure}\centering
\includegraphics[width=0.49\textwidth]{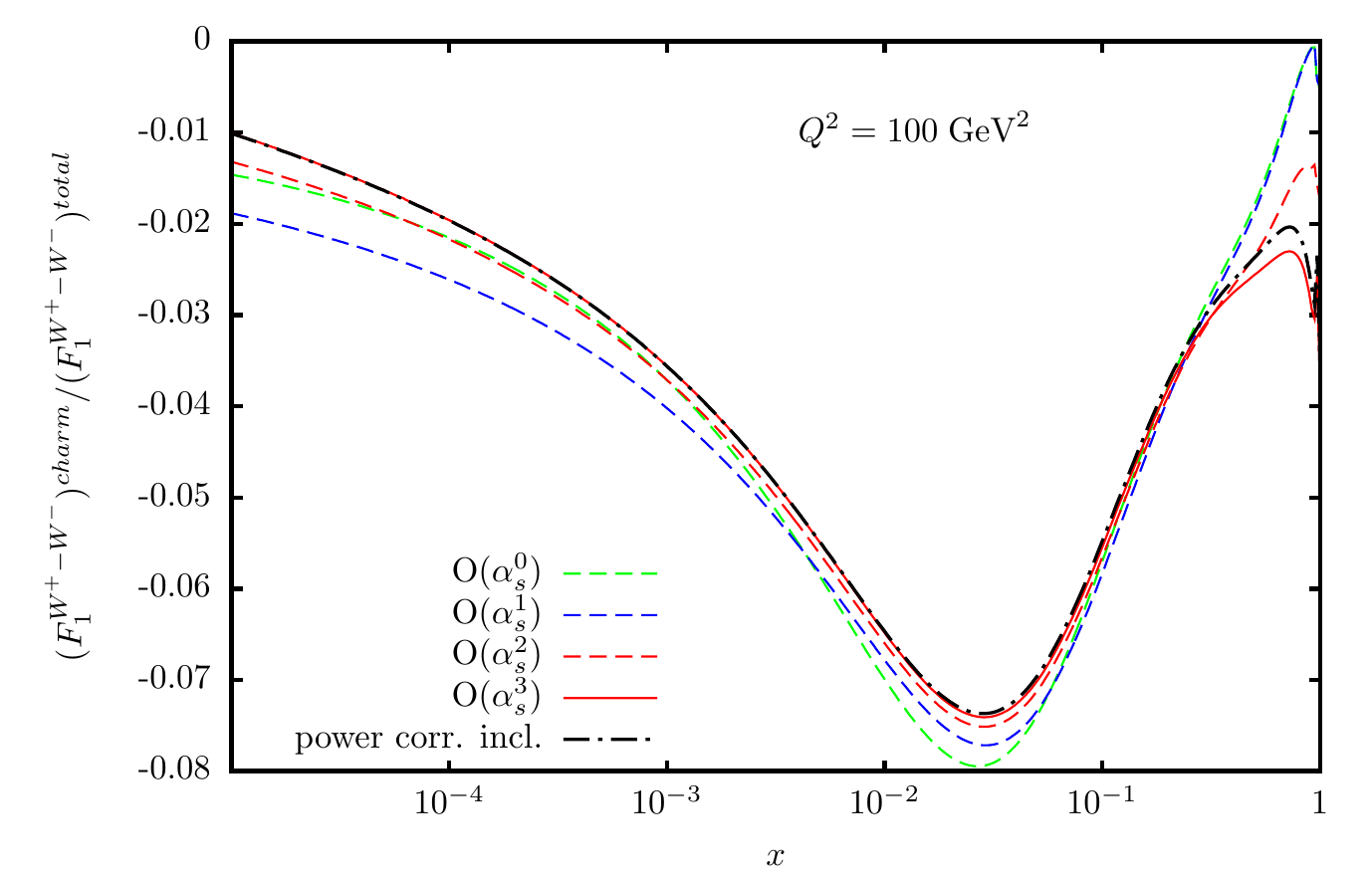}
\includegraphics[width=0.49\textwidth]{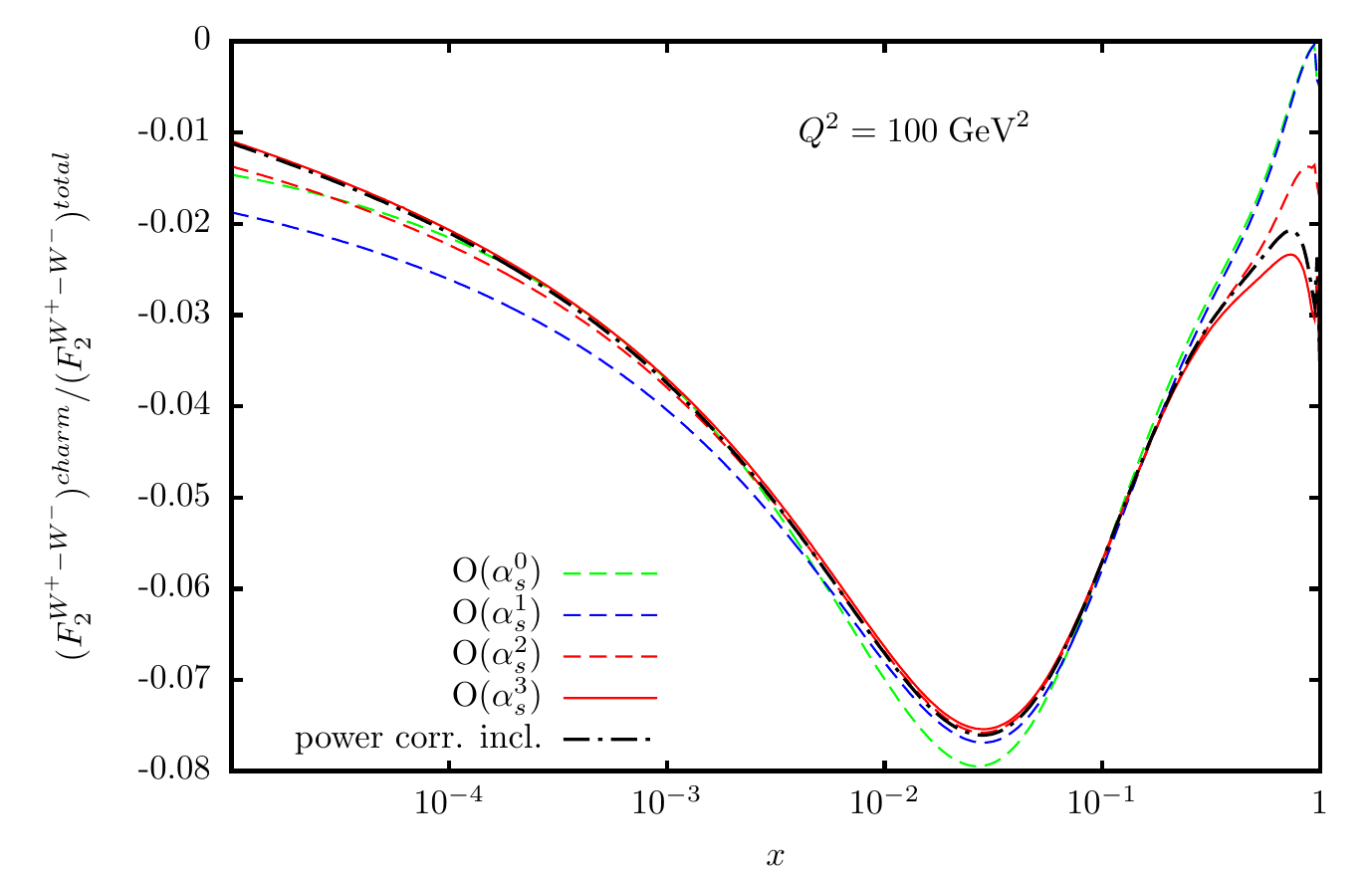}
\caption[]{The relative 3-loop charm quark corrections to the charged current structure functions
$F_{1(2)}^{W^+ - W^-}(x,Q^2)$ at $Q^2 = 100~\GeV^2$ for $m^{\rm pole} = 1.59~\GeV$ using the PDFs
of Ref.~\cite{Alekhin:2013nda}; form \cite{F1F2}.
\label{fig:3}}
\end{figure}
%-------------------------------------------------------------------------------------------------

\vspace*{-3mm}
\noindent
%-------------------------------------------------------------------------------------------------
\begin{figure}\centering
\includegraphics[width=0.49\textwidth]{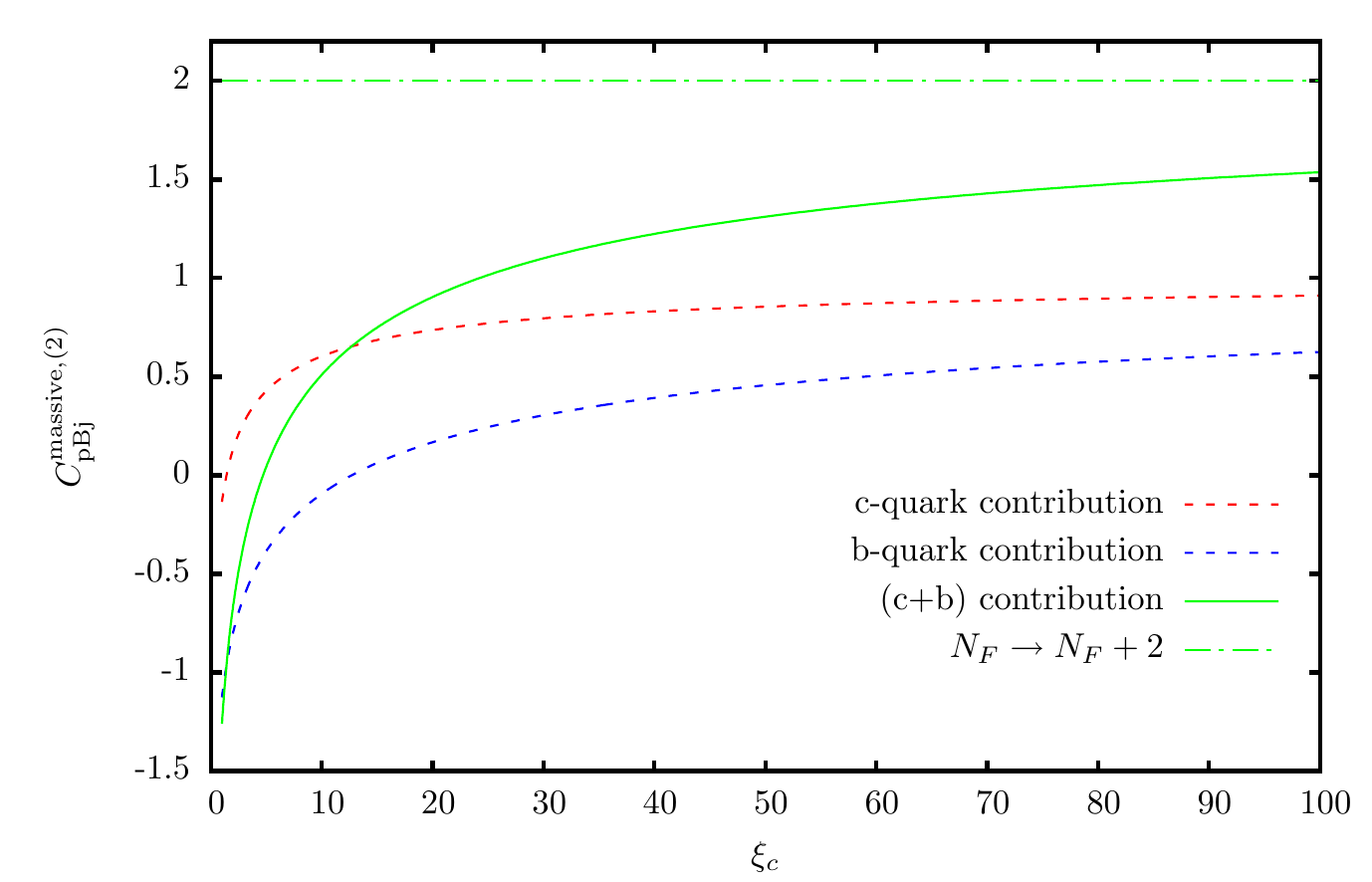}
\includegraphics[width=0.49\textwidth]{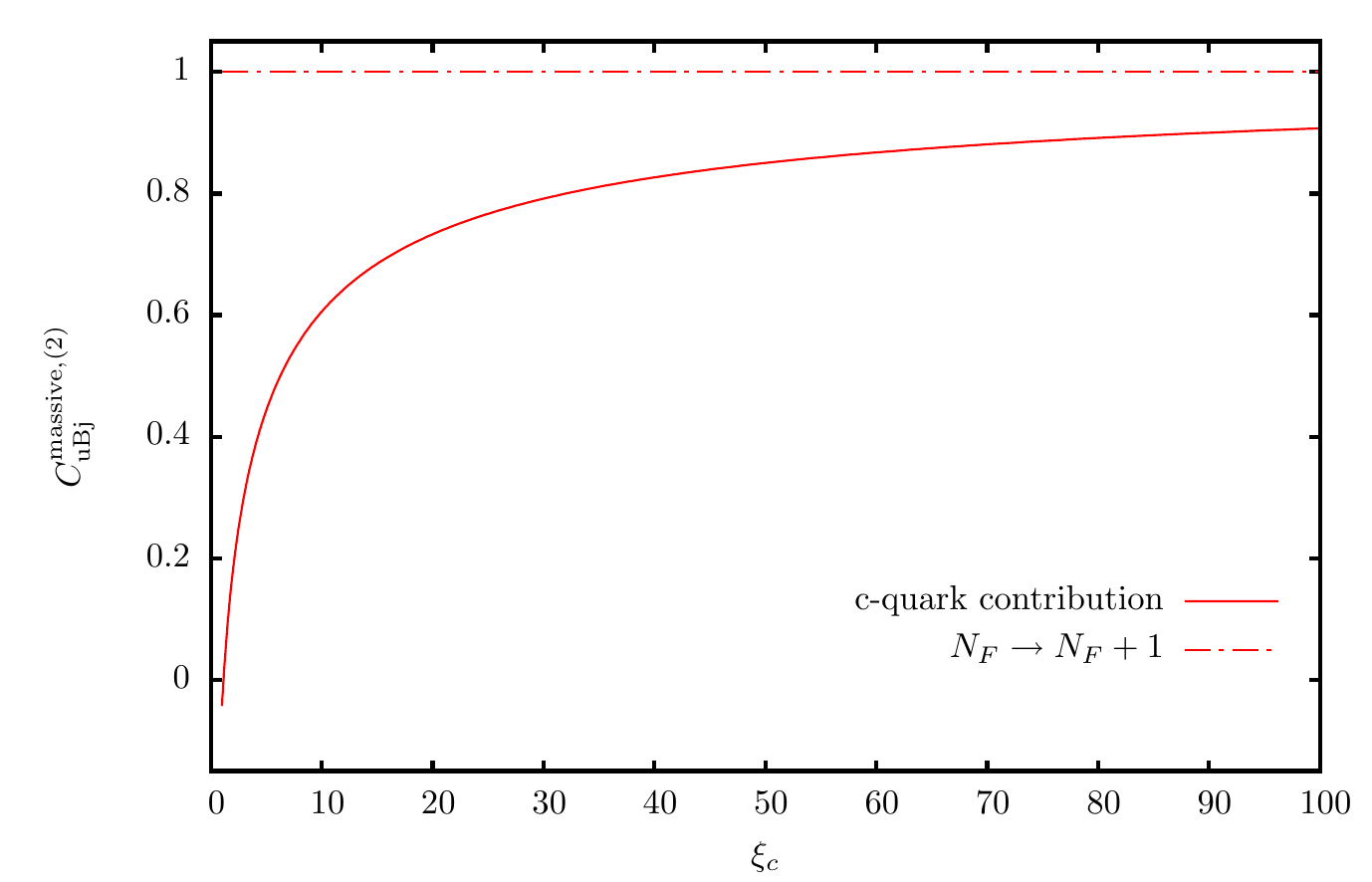}
\caption[]{Left panel:~The charm and bottom quark contributions to the polarized Bjorken sum rule as a function of 
$\xi_c = Q^2/m_c^2$, for $m_c^{\rm pole} = 1.59~\GeV$ and $m_b^{\rm pole} = 4.78~\GeV$ \cite{Agashe:2014kda}. 
The function $C_{\rm pBj}$ describes
the flavor excitation, with $C_{\rm pBj} = 1$ one heavy quark being effectively massless. Right panel:~The 2-loop 
charm quark contribution to the unpolarized Bjorken sum rule; from \cite{Blumlein:2016xcy}.
\label{fig:4}}
\end{figure}
%-------------------------------------------------------------------------------------------------

\vspace*{-3mm}
In the tagged flavor case a similar analysis was performed in Ref.~\cite{Blumlein:1998sh}\footnote{For the 
$O(\alpha_s^2)$ results on the tagged contributions to the structure functions see 
\cite{Buza:1995ie,Bierenbaum:2007qe}.}. It turns out that the tagged 
flavor case does not lead to a stable description at large scales $Q^2$, containing logarithmic terms, which are absent 
in the inclusive case due to heavy flavor loop effects in diagrams with massless final states. In the inclusive case
the decoupling of heavy flavors is uniquely described. Going to very low scales $Q^2$ for deep-inelastic scattering
logarithmic terms remain. However, the deep-inelastic description is limited to virtualities $Q^2$ of a certain size,
say $Q^2 \gsim Q_0^2 = 5~\GeV^2$ and cannot be applied below.

We show the effect of these contributions on the structure functions $F_{1(2)}$ in Figure~\ref{fig:3} (dash-dotted lines)
replacing the contributions up to 2-loops by the complete results. At $Q^2 = 100~\GeV^2$ the corrections do widely agree
with the asymptotic result but a small difference in the large $x$ range still remains.

One also obtains interesting results on different sum rules for the structure functions, such as 
the Adler- \cite{Adler:1965ty}, unpolarized Bjorken- \cite{Bjorken:1967px}, polarized Bjorken- \cite{Bjorken:1969mm}, 
and Gross-Lewellyn Smith sum rule \cite{Gross:1969jf}. While due to fermion number conservation the contribution to these sum 
rules is free of mass effects in the asymptotic range, the last three sum rules receive heavy quark mass effects. We have shown 
that the Adler sum rule does not receive any correction, also no target mass corrections.

Figure~\ref{fig:4} demonstrates the transition of the massive quark contributions to massless ones 
for the massive 2-loop corrections of the polarized and unpolarized Bjorken sum-rule as a function of $Q^2$.
One should note that the respective corrections enter the $N_F$-terms of the sum rule only, cf.~\cite{Blumlein:2016xcy}.
The transition of this term to 1 only proceeds slowly both for charm and bottom quarks as a function of $Q^2$. At lower 
scales $Q^2$ the corresponding contributions are even negative in case of the polarized Bjorken sum rule due to virtual 
corrections, which is of importance in experimental analyses.

\vspace*{-2mm}
%-------------------------------------------------------------------------------------------------
\section{3-Loop Two-Mass Corrections in the Asymptotic Region}
\label{sec:6}
%-------------------------------------------------------------------------------------------------

\vspace*{1mm}
\noindent
Starting at 3-loop order the massive Wilson coefficients contain Feynman diagrams in which two internal 
massive fermion lines are present leading to contributions of charm and bottom in individual terms, which 
cannot be separated. Due to this the variable flavor number scheme presented  in \cite{Buza:1996wv,Behring:2014eya}
is no longer applicable in the strict sense, but needs to be generalized. As $m_b^2/m_c^2 \sim 10$, one possibility
for large scales $Q^2 \gg m_c^2, m_b^2$ consists in decoupling both heavy quarks together. The corresponding
variable flavor number scheme has been worked out in Ref.~\cite{TWOM2} and requires to obtain as well the
two-mass contributions to the different OMEs at 3-loop order. Their moments $N= 2,4,6$ have already been calculated
\cite{TWOM2,TWOM1} projecting them to massive tadpoles and using the package {\tt Q2e} \cite{Q2E}. In the two-mass
case the renormalization of the OMEs, given in the single mass case in \cite{Bierenbaum:2009mv}, has to be extended.
Furthermore,
the 2-mass OMEs $A_{qq,Q}^{\rm NS},~A_{qq,Q}^{\rm NS,TR}$ and $A_{gq,Q}$ have been calculated for general mass 
assignment analytically and the calculation of $A_{Qq}^{\rm PS}$ is underway. In case of the OME $A_{gg,Q}$
all scalar topologies have been calculated. Here new classes of iterative integrals do emerge, which can be handled
with the package {\tt HarmonicSums.m} by now. While an expansion in the mass ratio $m_c^2/m_b^2$ is possible 
for fixed moments, it turns out not to be possible for a general Mellin variable $N$, thus requiring 
the complete calculation.

\vspace*{-2mm}
%-------------------------------------------------------------------------------------------------
\section{Conclusions}
\label{sec:7}
%-------------------------------------------------------------------------------------------------

\vspace*{1mm}
\noindent
After having calculated a set of 3-loop moments for all massive OMEs contributing to the massive 3-loop 
Wilson coefficients in the asymptotic region $Q^2 \gg m^2$ in 2009 \cite{Bierenbaum:2009mv}, progress 
has been made in the calculation of the  general $N$ results of these Wilson coefficients. All logarithmic 
contributions are known \cite{Behring:2014eya} and the Wilson coefficients $L_{q,}^{\sf NS}, L_{q}^{\sf PS},
L_{g}^{\sf S}$ and $H_q^{\sf PS}$ have been calculated along with the OMEs appearing as the transition matrix 
elements in the variable flavor number scheme at 3-loop order, 
cf.~\cite{Ablinger:2010ty,Behring:2014eya,Ablinger:2014vwa,Ablinger:2014nga}.
In case of the charged current structure functions and the non-singlet  contribution to the polarized structure functions
$g_{1(2)}(x,Q^2)$ a series of 3-loop Wilson coefficients have been calculated \cite{Behring:2015zaa,Behring:2015roa,F1F2}, 
completing the 2-loop charged current programme \cite{CC,Blumlein:2014fqa} before. Recently, the non-singlet $O(\alpha_s^2)$
corrections have been completed for the most important inclusive polarized and unpolarized neutral and charged current structure 
functions and the sum rules associated to them in the whole range $Q^2$ relevant for deep-inelastic scattering \cite{Blumlein:2016xcy}. 
It has been shown that the process of a heavy 
quark becoming effectively massless at high scales proceeds slowly, which should be considered in the matching in the 
different schemes being used in data analysis currently. Clearly, the matching at $m_Q^2 = Q^2$ is unfortunate since a heavy quark's 
velocity is not ultra-relativistic there.

At 3-loop order two-heavy-mass Feynman diagrams contribute to the OMEs. This requires a different renormalization compared to that of
the single-heavy-mass case and leads to a change of the associated variable flavor number scheme since these terms
are neither charm nor bottom contributions. One may, however, design a variable flavor number scheme in decoupling both contributions
together. Beyond a series of moments also the general $N$ contributions have been computed in some cases \cite{TWOM2}.

We have devised an algorithm to calculate massive Feynman diagrams containing local operator insertions, mapping the 
differential equations obtained from the IBP-relations into systems of difference equations, which are solved automatically
\cite{Ablinger:2015tua,Ablinger:2016yjz}. This algorithm factorizes the corresponding problem to first order structures as far as 
possible and into 
potential remaining terms, which cannot be factorized neither in $N$- nor $x$-space. If the latter terms are not present,
we receive an iterative sum or integral solution of the corresponding physical problem starting from differential equations in whatever 
basis and obtaining the emerging letter-representation of the contributing alphabet in an automatic manner together 
with a proof 
certificate. In this way we automatically detect non-factorizing contributions uniquely. They need a further separate 
treatment to obtain the solution.

In the present project various new mathematical structures have been found for Feynman-integrals in general.
While at the time of 1998 the harmonic sums yielded a sufficient systematic representation for massless 2-loop 
problems \cite{Blumlein:1998if,Vermaseren:1998uu}, the present massive 3-loop problems require Kummer-iterated 
integrals \cite{Moch:2001zr,Ablinger:2013cf}, cyclotomic iterated integrals \cite{Ablinger:2011te}, root-valued iterated 
integrals \cite{Ablinger:2014bra}, and further generalizations to elliptic and $_2F_1$-valued non-iterative letters in 
otherwise iterative integrals \cite{ELLIPT}. This list of structures will be growing further addressing processes at
even higher loops and with more scales. The packages {\tt Sigma, EvaluateMultiSums, SumProduction, RhoSum}, {\tt 
HarmonicSums} and {\tt MultiIntegrate}
\cite{SIG1,SIG2,EMSSP,RHOSUM,DECOUP,Ablinger:PhDThesis,Harmonicsums,Ablinger:2013cf,Ablinger:2011te,Ablinger:2014bra} 
were significantly extended or newly created in the project and are already in partly use in a series of other projects.

Presently we work on the completion of $A_{Qg}^{(3)}$, in which new mathematical structures appear, the solution of which has to be 
automated. 

\vspace*{-2mm}
%-------------------------------------------------------------------------------------------------

%-------------------------------------------------------------------------------------------------
\end{document}